\RequirePackage{snapshot}
\documentclass[12pt, draftclsnofoot, onecolumn]{IEEEtran}

\usepackage{verbatim}
\usepackage{fancyvrb} 

\usepackage{amssymb}
\usepackage{amsfonts}
\usepackage{amsmath,amssymb}%
\usepackage[usenames,dvipsnames]{color}

\ifCLASSINFOpdf 
\usepackage[pdftex]{graphicx} 
\usepackage[pdftex,colorlinks,
           bookmarks=true,%
           pdftitle=Wireless\ networks\ appear\ Poissonian\ due\ to \  strong shadowing,%
           pdfauthor=B.\ Blaszczyszyn\ -\ M.\ K.\  Karray\ -\  H.\ P.\ Keeler]{hyperref}  
\usepackage[numbers,sort&compress]{natbib} 
\usepackage{hypernat} 
\else

\usepackage[dvips]{graphicx} 
\usepackage[dvips,colorlinks,
           bookmarks=true,%
           pdftitle=Wireless\ networks\ appear\ Poissonian\ due\ to \  strong shadowing,%
           pdfauthor=B.\ Blaszczyszyn\ -\ M.\ K.\  Karray\ -\  H.\ P.\ Keeler]{hyperref}  
 \usepackage[numbers,sort&compress]{natbib} 
\fi 
\hypersetup{
   bookmarksnumbered,
   pdfstartview={FitH},
   citecolor={blue},
   linkcolor={red},
   urlcolor=[rgb]{0,0.55,0},
   pdfpagemode={UseOutlines}}

\graphicspath{{./}{./Figures/}} 

\newcommand{\OneOrTwoColumnDisplay}[2]{#1} 

\renewcommand\textcolor[2]{#2}  

\newcommand{\E}{\mathbf{E}}

\newcommand{\calR}{\mathcal{R}}

\newcommand{\Prob}{\mathbf{P}}
\newcommand{\Ind}{\mathbf{1}}
\newcommand{\R}{\mathbb{R}}

\newcommand{\dB}{\mathrm{dB}}
\newcommand{\logsd}{\sigma_{\dB}}

\newcommand{\calT}{\mathcal{T}}

\newcommand{\markTheta}{\widetilde{\Theta}}
\newcommand{\markN}{\widetilde{\Theta}}

\newcommand{\ir}{\mathbb{R}}

\newcommand{\ind}{\mathbf{1}}

\newcommand{\red}[1]{\textcolor{red}{#1}}

\newtheorem{thm}{Theorem}
\newenvironment{theorem}{\bf\begin{thm}\rm\em}{\end{thm}} 
\newtheorem{cor}[thm]{Corollary}
\newenvironment{corollary}{\bf\begin{cor}\rm\em}{\end{cor}} 
\newtheorem{lem}[thm]{Lemma}
\newenvironment{lemma}{\bf\begin{lem}\rm\em}{\end{lem}} 
\newtheorem{prop}[thm]{Proposition}
\newenvironment{proposition}{\bf\begin{prop}\rm\em}{\end{prop}} 
\newtheorem{rem}[thm]{Remark}
\newenvironment{remark}{\bf\begin{rem}\rm}{\end{rem}} 

\newtheorem{Fact}[thm]{Fact}

\title{Wireless networks appear Poissonian due to strong shadowing
 }

 \author{{\bf B. B{\l}aszczyszyn}, {\bf    M. K. Karray}
and  {\bf H.P. Keeler}\thanks{B. B{\l}aszczyszyn and H.P.~Keeler are with Inria-ENS, 
23 Avenue d'Italie, 75214 Paris, France; email: Bartek.Blaszczyszyn@ens.fr}
\thanks{M.~K.~Karray is with  Orange Labs, 38/40 rue
  G\'{e}n\'{e}ral Leclerc,  92794  Issy-les-Moulineaux, France; email: 
mohamed.karray@orange.com}}

\begin{document}

\maketitle
\thispagestyle{empty}
\begin{abstract}
Geographic locations of cellular base stations sometimes can be well fitted with spatial homogeneous Poisson point processes.  In this paper we make a complementary observation: In the presence of the log-normal shadowing of sufficiently high variance, the statistics of the propagation loss of a single user with respect to different network stations are invariant with respect to their geographic positioning, whether regular or not, for a wide class of empirically homogeneous networks.  Even in perfectly hexagonal case they appear as though they were realized in a Poisson network model, i.e., form an inhomogeneous Poisson point process on the positive half-line with a power-law density characterized by the path-loss exponent.  At the same time, the conditional distances to the corresponding base stations, \textcolor{blue}{given their observed propagation losses}, become independent and log-normally distributed, which can be seen as a decoupling between the real and model geometry.  The result applies also to Suzuki (Rayleigh-log-normal) propagation model.  We use Kolmogorov-Smirnov test to empirically study the quality of the Poisson approximation and use it to build a linear-regression method for the statistical estimation of the value of the path-loss exponent.
\end{abstract}


\let\thefootnote\relax\footnotetext{Separate parts of this paper were presented at WiOpt 2012, Paderborn, Germany\cite{blaszczyszyn2012linear} and Infocom 2013, Turin, Italy~\cite{hextopoi}. }

\newcommand{\thefootnote}{\arabic{footnote}}
\section{Introduction}

The immense increase of user-traffic is driving the need for more dense cellular networks and suitable analytic evaluation methods. 
The irregular positioning of base stations  deployed in dense urban areas implies that they are best assumed to be random,
which has motivated stochastic geometry models.
Base station positions often can be well fitted with homogeneous
Poisson point processes (cf~\cite{lee2013stochastic,guo2013spatial,typcell}), which enables or considerably simplifies analytic evaluation methods. 
``Worst-case'' arguments are also used to justify the use of Poisson models.
In this paper we \textcolor{blue}{revisit} an alternative argument, \textcolor{blue}{already considered in~\cite{hextopoi}}, based on the presence
of a shadowing fitted with the log-normal distribution of sufficiently large variance. 

More specifically, we \textcolor{blue}{revisit} the convergence result that a
broad range of {\em empirically homogeneous} network configurations
(for example, deterministic lattices or arbitrary random stationary
point patterns) give results appearing to the typical user as though the placement of base stations is a Poisson process when sufficiently large log-normal shadowing is incorporated into models with power-law path-loss functions\footnote{The power-law may be modified to remove its singularity at~0.}.
The rigorous statement involves the values of the propagation loss of a
single user with respect to all  base stations, called here the  propagation process.
When the variance of
the log-normal shadowing, assumed independent across different base stations,
tends to infinity, these values (considered as a point process on the
positive half-line),  appropriately rescaled,  converge to an inhomogeneous Poisson point process
with a power-law density 
characterized  (up to a multiplicative constant) by the path-loss exponent.
This is exactly how a Poisson network ``appears'' to its typical user.

\textcolor{blue}{In this paper we present also two extensions of the above convergence result. 
Firstly, we consider some  arbitrary
characteristics, for example, the type of the station in a $K$-tier
network as in~\cite{DHILLON2012}, which may depend on the shadowing.
We  characterize the distribution of these characteristics  in the asymptotic regime.
We study also the  geographic distances to the bases stations (from the
given user measuring the propagation loss values) and find that 
the conditional distances to the base stations given
their propagation loss values 
asymptotically become independent and log-normally distributed.
This is also how a Poisson network with  log-normal shadowing
``appears'' to its typical user. However, in the limiting regime 
(unlike in the Poisson network) these distances are identically
distributed, which means that the value of the measured propagation loss
(small or large) does not carry any information about the distance
of the corresponding station. This can be seen  as some kind of
decoupling between the real and model geometry and sheds more light on the limitations 
of the applicability of the Poisson model 
whenever its only justification is a strong shadowing. In this latter case,  the  network characteristics
entirely based on the values of the propagation losses can be reliably approximated by the corresponding
functionals of the Poisson process, which does not however represent  the Euclidean  geometry of the network.
}

 The above convergence results also apply to  composite
 fading-shadowing models when they consist of a product of two (or
 more) independent random variables with one being
 log-normal. Examples of such models include Suzuki (or
 Rayleigh-log-normal)~\cite{suzuki1977statistical} shadowing, a
 generalization of it~\cite{withers2012generalized}, and
 Nakagami-log-normal shadowing. For more details on these models and a
 comparison of their statistical estimators, see Reig and
 Rubio~\cite{reig2013estimation}.

\textcolor{blue}{The presentation of the  convergence results is preceded with new results regarding the
invariance of the marked Poisson network model, which help to understand the asymptotic scaling of the general marked
model.}

Asymptotic results are useful in practice whenever the convergence is
fast enough for the limiting object to be a reasonable approximation
of the actual (pre-limit) situation. \textcolor{blue}{A precise analysis of the speed of
this convergence is beyond the scope of this paper}, \red{and would arguably require a different proof technique.} However,
\textcolor{blue}{revisiting~\cite{blaszczyszyn2012linear}}, 
we address this issue 
by comparing the empirical distribution function of the path-loss to the strongest base
station and of the signal-to-interference-and-noise ratio (SINR) 
simulated in perfectly hexagonal network and
measured in urban areas of real operational networks
to the corresponding analytic results available for Poisson model.
We show that the Poisson  model can fit both the statistics of the
perfect hexagonal and the ``real'' network.
To make this claim more quantitative, we use the 
Kolmogorov-Smirnov tests on the statistics of the hexagonal network
and show that it cannot be significantly distinguished by a single user
(measuring the propagation loss to the serving base station)
from a Poisson network in the presence of the log-normal shadowing of
the logarithmic standard deviation of about 10dB. This is a  realistic assumption
for outdoor and indoor wireless communications in many urban scenarios.

Having justified the Poisson approximation in a ``real'' network scenario,
we use this limiting model to build  a new statistical
method for estimating the exponent of the path-loss function 
based on propagation loss data collected by
users with respect to their serving base stations. 
This new method complements existing methods.
We illustrate the proposed method on both simulation 
and real-world data from a network operator in Europe.

{\em The remaining part of the paper} is organized as
follows. 
\textcolor{blue}{In Section~\ref{s.related-work} some related works are briefly discussed.} 
In Section~\ref{s.Poisson} we recall  the Poisson network model with its  useful results, some of which are extended. 
Our main Poisson convergence results are presented in Section~\ref{s.convergence}. Their proofs are deferred to the \hyperref[s.Appendix]{Appendix}. In Section~\ref{s.Regression} we numerically verify the quality of the asymptotic Poisson approximation 
and present a new  linear-regression method for the estimation of path-loss exponents.

\section{Related works}
\label{s.related-work}
The main convergence result proved in this paper can be rephrased as follows:
In the presence of the log-normal shadowing of sufficiently high variance,
the statistics of the propagation loss of a single user with respect to different
base stations are  essentially invariant with  respect to the
exact geographic positioning of the base stations, whether  regular or not,
for a wide class of empirically homogeneous networks. 
Even in perfectly hexagonal case they appear as though they were
realized in a Poisson network model.  Brown~\cite{brown2000cellular}
first observed  this by simulation, later confirmed independently by  B\l aszczyszyn and Karray~\cite{blaszczyszyn2012linear} who suggested using classical convergence results of random translations of point processes; see Daley and Vere-Jones~\cite[Section 11.4]{daleyPPII2008}. This approach was first adapted and applied in this setting by B\l aszczyszyn, Karray and Keeler~\cite{hextopoi}, which forms part of the results presented here. 

The fact that an arbitrary network can be approximated (from the point of view of a single user) by a Poisson model 
is very useful. This latter model has already been extensively studied. 
In particular, it enjoys the following very useful property, 
here referred to as {\em propagation invariance}, which stems from using a  power-law as the path-loss function:
The statistics of the propagation loss of a single user with respect to different base stations (which we called the propagation process~\footnote{Also called  ``path-loss process with fading'' in~\cite{HAENGGI2008}.}) depend on the fading distribution through only one moment. This property, provides considerable tractability and has led to the concept of so-called {\em equivalent networks} from the perspective of the typical user; cf~\cite{equivalence2013}, where extensions to heterogeneous networks 
are also presented.

This convenient property has been observed independently in physics
models~\cite{ruelle1987mathematical,panchenko2007guerra} and  network
models~\cite{madhusudhanan2009carrier,blaszczyszyn2010impact,vu2014analytical};
see~\cite{blaszczyszyn2014studying} and references therein for further
details. 
For example, compare~\cite[Corollary 10]{equivalence2013}
and~\cite[Lemma 2]{panchenko2007guerra}, which effectively both give
the same equivalence result for, in our setting, the marked
propagation process of the typical user.  In the physics context,
these invariance results imply Bolthausen-Sznitman invariance
property (cf~\cite[Eq.~(2.26)]{panchenko2013sherrington}) for the
Poisson-Dirichlet process, which is used  to study the
Sherrington-Kirkpatrick model for spin glasses (types of disordered
magnets). 
This latter process~\footnote{It should not be confused with the better known  Poisson-Dirichlet process of Kingman~\cite{KINGMAN:1993}. In fact 
both processes are special cases of two-parameter Poisson-Dirichlet process extensively studied in~\cite{pitman1997two}.}
 is exactly the so-called {\em signal-to-total-interference ratio} (STIR) process, which in turn is trivially related to the SIR process in (interference-limited) Poisson networks~\cite{blaszczyszyn2014studying,sinrtwopd2014}.

 Recently several characteristics related to the propagation process in the Poisson model have been studied. For example, closed and semi-closed expressions for the coverage probability in so-called multi-tier network models~\cite{ANDREWS2011,MADBROWN2012,MUKHERJEE2012,dhillon2011tractable,mukherjee2013analytical}, with the concept of $k$-coverage being later introduced~\cite{kcovsingle,blaszczyszyn2014studying}. More recently, these models have been advanced by studying the coverage probability under signal coordination~\cite{giovanidis2013stochastic,tanbourgi2014analysis} and interference cancellation schemes~\cite{zhangsuccessive2013,wildemeersch2013successive}, with a recent contribution being the joint probability density of the order statistics of the process  formed from the SINR values of the typical user~\cite{blaszczyszyn2014studying}.  
The recently observed relation between the SINR values and a type of Poisson-Dirichlet process (cf~\cite{sinrtwopd2014}) can potentially bring some further progress to this subject.

\textcolor{blue}{Finally, as we have already explained, the asymptotic Poisson model
does not represent well the ``real'' geometric locations of base
stations. Several papers propose and study ``more realistic'' geometric
models of the network based on, for example, determinantal point processes~\cite{miyoshi2014cellular,nakata2014spatial,li2014statistical,Deng2015}, \red{of which the Poisson process is a special case}.}

\section{Poisson network model}
\label{s.Poisson}
The goal of this section is to present the Poisson network model which
will be proved in the next section to be a limit of an arbitrary,
stationary network model subject to strong log-normal shadowing.

 On $\R^2$, we model the base stations with a homogeneous Poisson point process
 $\Phi=\{X_i\}$ with density $\lambda$.  We take the ``typical user''
 model approach where one assumes a typical user is located at the
 origin and consider what he perceives or experiences in the network.
 Given $\Phi$, let $\{S_i\}$ be
 independent and identically distributed (iid) positive random
 variables that represent the propagation effects (shadowing and/or
 fading) experienced by the typical user with respect to the
 respective stations.  Define the path-loss function as
\begin{equation}\label{e.DistanceLoss}
\ell(x)=(K|x|)^{\beta},
\end{equation}
with path-loss constant $K>0$ and path-loss exponent $\beta>2$.

We define the {\em propagation process}, considered as a point process on the positive half-line $\mathbb{R}^{+}$, as
\begin{align}
\Theta:&= \left\{\frac{\ell(X_i)}{ S_i } : X_i\in \Phi \right\} \\
&= \{L_i\}\,
\end{align}
where $L_i$ is called {\em propagation loss} from station $X_i$.
It has been observed that the  propagation process exhibits a  convenient invariance
property~\cite{ruelle1987mathematical,madhusudhanan2009carrier,blaszczyszyn2010impact,vu2014analytical}.~\footnote{
\label{foot:loss-or-gain}
\textcolor{blue}{It is \red{somewhat} more convenient and natural to consider the propagation-loss
process $\Theta$, which does not have infinitely many small values.
This approach is compliant with the existing engineering literature
on this subject. However, mathematically (using an appropriate
topological formalism) it is possible to consider
the process of the received powers; cf e.g.~\cite{sinrtwopd2014}.}}

\begin{lemma}[Propagation invariance]\label{l.invariance}
Assume that
\begin{equation}\label{momcond}
  \E(S^{\frac{2}{\beta}})  <\infty .
\end{equation}
Then the propagation process $\Theta$ is an inhomogeneous Poisson point
process with intensity measure
$\Lambda\left(  \left[  0,y\right)  \right)  =a y^{\frac{2}{\beta}}$,
where the propagation constant is
\begin{equation}\label{e.a}
a:=\frac{\lambda\pi
  \E(S^{\frac{2}{\beta}})}{K^{2}}\,.
\end{equation}
\end{lemma}
This is a  well-known result. In this paper we will prove
its extension, Proposition~\ref{p.markedinvariance}. All our proofs 
are  given in the~\hyperref[s.Appendix]{Appendix}.

\begin{remark} 
\label{r:truncated}
The above result allows one to represent random fading and/or
shadowing (called {\em propagation effects} in what follows) by setting, for example,  $S=1$ and replacing $\lambda$ with
$\tilde \lambda=\lambda\E (S^{2/\beta})$ or (equivalently) replacing
$K$ with $\tilde K=K/\sqrt{\E(S^{2/\beta})}$.
This invariance with respect to the distribution of the propagation
effects $S$ can be rephrased by saying that the Poisson network model
is a fixed point of a network transformation which consists of
perturbing path-losses from all base stations to the typical user by
some random (iid) propagation effects. We will see in the next section
that this fixed point is also a limit of an arbitrary network subject
to ``strong'' perturbations of the same kind.
\end{remark}

In what follows we extend the above result by considering distances
and some additional marking of the base stations in conjunction with
the propagation process.

Given $\Phi$, denote by $R_i=|X_i|$ the distance from base station $X_i$
to the typical user. Moreover for each $X_i\in \Phi$, let $T_i$ be  
some additional parameter or a vector  of parameters of 
 the base station $X_i$ with values in some state space
 $\calT$.~\footnote{For example, in a multi-tier  network composed of
 $K$ types of stations, $T_i\in\calT=\{1,\ldots,K\}$ can be the type
   of the station. One can also model different  SINR thresholds of the
   base stations assuming  $T_i\in\ir^+$, which leads to a
 generalization of the multi-tier model called a random heterogeneous
 cellular network~\cite{equivalence2013}.} We will call $T_i$  the {\em type} of
    base station $X_i$. While $R_i$ clearly
 depends on $X_i$, we assume that $T_i$ does not depend on the base
 station location but  may depend on the propagation effects $S_i$
 from this base station. More specifically, we assume that given
 $\Phi$, $(S_i,T_i)$ are iid across $i$. Denote by
   $(S,T)$ a generic random variable of this distribution and by
   $G^T(\tau)=\Prob(T\in\tau)$, and
   $G^{T|S}(\tau|s)=\Prob(T\in\tau|S=s)$ for
 $\tau\subset\calT$, the marginal distribution on $T$ and its
   conditional distribution given $S=s$, respectively.
Thus 
\begin{equation}\label{e.Theta}
\markTheta \equiv\{(L_i,(R_i,T_i))\}\,.
\end{equation}
forms on $\R^+ \times \R^+\times \cal T$ an independently marked point process.

In essence what the typical user ``perceives'' in our network model is
represented by  $\markTheta$, hence its distribution determines all
the characteristics of the typical user that can be expressed in terms
of its propagation losses (for example,  SINR, spectral and energy
efficiency, etc).  The process $\markTheta$ is also a Poisson point
process by the next result, which can be seen as an extension  of a previous result ~\cite[Lemma 1]{equivalence2013}.

\begin{proposition}\label{p.markedinvariance}
Under the assumptions of Lemma~\ref{l.invariance}
the propagation process $\markTheta$ is an independently marked inhomogeneous Poisson point process  with intensity measure
\begin{align}
\tilde\Lambda((0,y)\times(0,\rho)\times\tau)&:=\E[\{i: L_i\le y, R_i\le \rho, T_i\in\tau\}]\nonumber\\
&= \int_0^{y} G_u (\rho,\tau)\Lambda(du)  \label{e.tLambda},
\end{align}
where $y,\rho\ge0$, $\tau\subset \cal{T}$
and 
\begin{align}
G_u (\rho,\tau)&= \frac{\E [S^{2/\beta} \Ind (S
  \leq (K\rho)^\beta/u) \Ind(T \in \tau)  ]}{\E(S^{2/\beta})}\,.
\label{e.Gu}
\end{align}

\end{proposition}
The proof is given in Appendix~\ref{app.markedinvariance}.

\begin{remark}
Note that $G_u$ represents the joint conditional  distribution of the
distance to a base station and its type given its propagation loss with
respect to the  typical user is equal  $u$ 
\begin{align}
G_u (\rho,\tau)&=\Prob (R\leq \rho , T\in \tau |L=u).
\end{align}
Note that, while the conditional distribution of $R$ given $L=u$
indeed depends on $u$
\begin{align}
G^R_u (\rho):=G_u(\rho,{\cal T})= \frac{\E [S^{2/\beta} \Ind (S
  \leq (K\rho)^\beta/u)]}{\E(S^{2/\beta})}\,,
\label{e.Gurho}
\end{align}
this is not the case for the conditional distribution of the base
station type
\begin{align}
\tilde G^T(\tau):=G_u(\infty,\tau)= \frac{\E [S^{2/\beta}\Ind(T \in \tau)  ]}{\E(S^{2/\beta})}\,.
\label{e.Gut}
\end{align}  
In other words, the {\em  type of the base station is independent of the
propagation loss with which it is received by the typical user}, despite the
fact that it might depend on the respective propagation effects
(recall that we allow dependence between the  components of the
vector $(S,T)$).  
\end{remark}
 
Note that, in general,  the distribution of the type of the base station, given
its propagation loss known to the user, $\tilde G^T(\tau)$ expressed by~(\ref{e.Gut}) is different from the
distribution of this type, given the location of the base station,
$G^T(\tau)=\E[\ind(T\in\tau)]$. In fact, the former is a $S^{2/\beta}$-biased
modification of the latter. 

An interesting observation regarding the 
log-normal distribution of propagation effects $S$ (this distribution  will play a special
role in the next section) is that it implies that the conditional
distance to base stations is also log-normal.
More specifically, for $\mu\in\ir$ and $\sigma>0$ let 
\begin{equation}\label{e.lognormalRV}
S=\exp(\mu+\sigma Z),
\end{equation}
where $Z$ is the  standard Gaussian or normal random variable (with zero mean
and unit variance) whose distribution will be denoted by $G_Z$. 
This parametrization of the log-normal $S$ implies  
\begin{equation}\label{e.S2/b}
\E(S^{2/\beta})
=\exp\left[\frac{2\mu}{\beta}+\frac{2\sigma^2}{\beta^2}\right]\,.
\end{equation}

\begin{proposition}\label{p.LNGuT}
Assume log-normal propagation effects $S$ as in~(\ref{e.lognormalRV}).
Then
\begin{equation}\label{e.Gu-LN}
G_u (\rho,\tau)= \E\Bigl[\Ind\Bigl(Se^{2\sigma^2/\beta}  \leq (K\rho)^\beta/u\Bigr)
G^{T|S}(\tau|Se^{2\sigma^2/\beta})\Bigr]\,
\end{equation}
for $\tau\subset \calT$, where $G^{T|S}(\tau|s)$ is the conditional
distribution of $T$ given $S$. In particular, 
\begin{align}
\tilde G^T(\tau)&=\E[G^{T|S}(\tau|Se^{2\sigma^2/\beta})], \label{e.GT} \\
 G^R_u(\rho)&=  \E[\Ind (Se^{2\sigma^2/\beta}   \leq (K\rho)^\beta/u)] \label{e.GR} .
\end{align}
\end{proposition}
The proof is given in Appendix~\ref{app.LNGuT}.

\begin{remark}\label{r.LN-distance}
Proposition~\ref{p.LNGuT}
says that the distribution of the distance to \textcolor{blue}{ a
  random}  base station,  given its propagation loss $u$, is a log-normal random
  variable,   specifically having the distribution of 
\begin{equation}\label{e.RPois}
R_u:=\frac{u^{1/\beta}}{K}\exp\Bigl[\frac{2\sigma^2}{\beta^2}\Bigr]S^{1/\beta}\,.
\end{equation}
\end{remark}

\section{Convergence of an arbitrary network to the Poisson one under log-normal shadowing} \label{s.convergence}
In this section we derive a useful convergence result rigorously showing that
the infinite Poisson model can be used to analyze the characteristics
of the typical user in the context of any {\em fixed (deterministic!)
placement of base stations}, meeting some \emph{empirical homogeneity}
condition,  provided there is {\em sufficiently strong log-normal shadowing}. 

Let  $\phi=\{X_{i}\}_{i\in \mathbb{N}}$ be a deterministic, locally finite  collection
of points (atoms) on $\R^2$ without an atom at the
origin.\footnote{Such an atom, together with the power-law path-loss
function, would give a  ``fixed'' atom at the origin in the propagation
process for any realization of the shadowing.} We think of them as representing location of
base stations in a ``real'' network.  Let $B_0(r)=\{x\in\R^2:|x|<r\}$
a ball of radius $r$, centered at the origin. We require the following
{\em empirical homogeneity condition} for $\phi$:
there is a constant $0 < \lambda < \infty$, such that as $r\rightarrow \infty$ 
\begin{equation}\label{lambdaDef}
 \frac{\phi(B_0(r))}{\pi r^2} \rightarrow \lambda.
\end{equation}
This general condition is satisfied by a wide class of base station
configurations including any lattice  pattern (with or without each
point independently perturbed) and almost any realization of an
arbitrary stationary, ergodic point process.

We present a convergence result demonstrating that a Poisson model can be used to study the functions or performance characteristics of the propagation processes and the base station distances from the view of the typical user, provided the placement of base stations meets the \emph{empirical homogeneity} condition (\ref{lambdaDef}) and there is  {\em sufficiently strong log-normal(-type) shadowing}. 

Let the  shadowing  $S_i=S_i^{(\sigma)}$ between the station $X_i\in\phi$ and the
origin  be iid (across $i$) log-normal random variables parametrized
as in~(\ref{e.lognormalRV}), with $\mu=-\sigma^2/2$. This latter
assumption makes $\E[S]=1$.\footnote{The assumption $\E[S]=1$ is a
  matter of convention, which is, for example, adopted in the  COST
  Walfisch-Ikegami model,  cf.~\cite[\S 2.1.6 and \S
    4.4.1]{Cost231_1999}. Another option, for example,  in~\cite{Jakes1974},
  is to assume that $S$ expressed in dB is centered:
$\E[10\log_{10}S]=0$, which is  equivalent to our model 
with the constant $K$ replaced by $e^{\sigma^2/(2\beta)}K$.}

We will study the propagation process generated by the deterministic
point pattern $\phi$ in the presence of the  log-normal shadowing as
$\sigma$ increases to infinity. In order to 
obtain a non-trivial limit,  we need to rescale 
the propagation loss process, which, in light of~(\ref{e.a}), we achieve by multiplying the path-loss
constant  $K$ by $\sqrt{\E(S^{2/\beta})}$, that is, by considering 
\begin{equation}\label{e.Ksigma}
K^{(\sigma)}=K \exp\left[\frac{\sigma^2(2-\beta)}{2\beta^2}\right],
\end{equation}
where  $K>0$ and $\beta>2$.

As in Section~\ref{s.Poisson}, we consider the point process on $\mathbb{R}^+$
of  propagation
  losses  experienced by the user located at the origin with respect to the stations
in~$\phi$
$$
 \Theta^{(\sigma)}:=\left\{\frac{ {K^{(\sigma)}}^{\beta} |X_i|^{\beta}
 }{S_i^{(\sigma)}}: X_i\in\phi\right\}=\{L^\phi_i \}\,.
$$
We consider also the analogous process of  propagation
  losses  
\begin{equation}
 \bar\Theta^{(\sigma)}:=
\{L^\phi_i :  a_\sigma < |X_i| <b_\sigma\}\,
\end{equation}
where the stations in $\phi$ that are closer than $a_\sigma$ and farther
than $b_\sigma$ are ignored, for 
$0\le a_\sigma< b_\sigma \le \infty$ satisfying
\begin{align}\label{a_n}
\frac{\log( \max(a_\sigma,1))}{\sigma^2} &\rightarrow 0,\\
\label{b_n}
 \frac{\log(b_\sigma)}{\sigma^2} &\rightarrow \infty.
\end{align}
\textcolor{blue}{(For the simplicity of notation we keep the dependence of  $\bar\Theta^{(\sigma)}$ on $a$ and $b$
  implicit.)
The reason for considering a truncated  pattern will be  clear in view of the second statement of
Remark~\ref{r:truncated}. The specific values of $a$ and $b$ follow form the proof.}

We present now our first convergence result regarding the non-marked
propagation processes $\Theta^{(\sigma)}$ and $\bar\Theta^{(\sigma)}$.
\textcolor{blue}{As we shall see, both processes  
have the same Poisson limit.}

\begin{theorem}\label{mainResult}
Assume  homogeneity condition (\ref{lambdaDef}). Then
$\Theta^{(\sigma)}$
converges weakly as $\sigma\rightarrow \infty$ to the
 Poisson point process on $\R^+$ with the intensity measure $\Lambda$
specified in Lemma~\ref{l.invariance} with $a=\lambda\pi/K^{2}$. 
Moreover, $\bar\Theta^{(\sigma)}$
also converges weakly $(\sigma\rightarrow \infty)$ 
to the Poisson point process with the same intensity measure,
provided conditions~(\ref{a_n}) and~(\ref{b_n}) are satisfied.
\end{theorem}

The  proof of Theorem~\ref{mainResult} is deferred to Appendix~\ref{AppMainTh}.

\begin{remark}\label{remark2}
The above result, in conjunction with Lemma~\ref{l.invariance}
 says that the {\em infinite Poisson model can be used
to approximate the characteristics of the typical user for a very general
class of homogeneous patterns of base stations}, including the standard
hexagonal one. The second statement of this result says that this
approximation  remains valid for {\em sufficiently large but finite patterns}.
\end{remark}

\begin{remark}
The path-loss model~(\ref{e.DistanceLoss}) suffers
from having a singularity at the origin. This issue is often
circumvented by some  appropriate modification of the path-loss function 
within a certain distance from the origin.  
The  second statement of Theorem~\ref{mainResult} with
$a_\sigma=\text{const}>0$  shows that such a modification is not significant
in the Poisson approximation.
\end{remark}

In what follows we will consider the pattern of base stations $\phi$ independently marked by the types in $\calT$ 
and we will be   interested in the propagation process generated by $\phi$, marked by these types and also the distances 
of the respective base station,  analogous to
$\markTheta$ in~(\ref{e.Theta}).  We have already
seen in the second statement of Theorem~\ref{mainResult} that the
\textcolor{blue}{conditional geographic distances to  base stations,
  whose  path-loss values are less than some given value}, in the limit of  $\sigma\to\infty$ escape to
infinity~\footnote{\label{foot:geometry}\textcolor{blue}{In other words, the base station received with relatively large signal
    under strong shadowing are typically not the closest ones. It is rather the
opposite: they might be  more and more far from the observer. (The non-degenerate Poisson limit
arises under  an appropriate rescaling of the pat-loss constant~\ref{e.Ksigma}.)  
In~(\ref{e.Rasymptotic}) we will give the asymptotic rate of this ``escape'' to infinity.}}.
 In order to study this behaviour 
we define the function
\begin{equation}\label{e.tRsigma}
  \calR(r):=\frac{\beta}{\sigma}\log r -\frac{\sigma}{\beta},\quad r\ge0\,
\end{equation}
and consider $R_i^{(\sigma)}:=\calR(|X_i|)$.
The above scaling can be deduced from the conditional distribution of
the distance given the propagation loss in the Poisson network, cf.~(\ref{e.RPois}).
Similarly, we have to rescale the conditional distribution of the marks, in case they depend on the shadowing $S^{(\sigma)}$.
To this regard, let $G^{T|Z}(\tau|z)$, $z\in\ir$, $\tau\in\calT$, be a given  probability kernel from $\ir$ to $\calT$~%
\footnote{For each $z\in\ir$, $G^{T|Z}(\cdot|z)$ is a probability measure on $\calT$.}
and we assume the following conditional distribution of the mark
$T_i^{(\sigma)}$ of the base station $X_i$ given the shadowing $S_i^{(\sigma)}$
\begin{equation}\label{e.GTsigma}
G^{T^{(\sigma)}|S^{(\sigma)}}(\tau|e^{-\sigma^2/2+\sigma z})=G^{T^{(\sigma)}|Z}(\tau|z)=G^{T|Z}(\tau| z-2\sigma/\beta)
\,.
\end{equation}
Both scalings can be  deduced from~(\ref{e.Gu-LN}) as explained in the Appendix~\ref{AppMainTh2}. 
Denote 
\begin{equation}\label{e.Grho}
G(\rho,\tau):=\int_{-\infty}^\rho G^{(T|Z)}(\tau|z)\, G_Z(dz)\,,
\end{equation}
where $G_Z$ is the distribution function  of $Z$ (standard Gaussian variable).

Consider an independently  marked propagation loss  process 
generated by fixed point pattern~$\phi$
\begin{align}\label{e.TildeThetaSigma}
 \markN^{(\sigma)}:&=\left\{\Bigl(L_i^\phi, (R_i^{(\sigma)},T_i^{(\sigma)})
 \Bigr)\right\}
\end{align}

We have the following refinement of Theorem~\ref{mainResult}.

\begin{theorem}\label{mainResult2}
Assume  homogeneity condition (\ref{lambdaDef}).
Consider the marked propagation  process 
$\markN^{(\sigma)}$ given by~(\ref{e.TildeThetaSigma}).
Then,  as $\sigma\rightarrow \infty$,  $\markN^{(\sigma)}$ converges
weakly to the iid 
 marked Poisson point process on $\R^+\times
\R\times\calT $ with intensity measure 
\begin{align}
\tilde\Lambda((0,y)\times(-\infty,\rho)\times\tau)
=G(\rho,\tau)\Lambda((0,y))  \label{e.tLambda-Conv},
\end{align}
where $\Lambda$ is as in Lemma~\ref{l.invariance} with
$a=\lambda\pi/K^{2}$  and $G(\rho,\tau)$ is given by~(\ref{e.Grho}).
\end{theorem}
The  proof of Theorem~\ref{mainResult2} is deferred to Appendix~\ref{AppMainTh2}.

\begin{remark}
 Note that $G(\rho,\tau)=\Pr(Z\le \rho,T\in\tau)$ represents the asymptotic (large $\sigma$) distribution of the
modified distance $R_i^{(\sigma)}$ to a base station and its type
$T_i^{(\sigma)}$. Regarding this distribution, we have the following observations:
\begin{enumerate}
\item This distribution, unlike in~(\ref{e.tLambda}), does not depend on the observed
value of the propagation loss --- these values form a Poisson process of
intensity $\Lambda$.  
This means that  asymptotically, for large $\sigma$, the distances  and type of the base station  whose
  propagation losses is observed by the typical user are independent of the registered value of this propagation loss. 
Consequently, the asymptotic (large $\sigma$) Poisson-point-process representation of
a real network is in fact {\em  decoupled} from the underlying ``real''
geometry. 
\item Express $R_i=|X_i|$ in terms of $R_i^{(\sigma)}$ (solving~(\ref{e.tRsigma})) and
consider the (limiting) standard Gaussian distribution for  $R_i^{(\sigma)}$  to conclude that for large $\sigma$ 
\begin{equation}\label{e.Rasymptotic}
R_i^{(\sigma)}\approx \exp\Bigl(\frac{\sigma^2}{\beta^2}+\frac{\sigma}{\beta}
Z\Bigr)\,.
\end{equation}
In other words, the
distance  $R_i$ is asymptotically log-normal. More precisely, 
note that the right-hand side of~(\ref{e.Rasymptotic}) is equal to 
$Ku^{-1/\beta}R_u$, where $R_u$ given in~(\ref{e.RPois})
is  the exact (not asymptotic) conditional
distance to the base station received with the propagation loss~$u$   
in the Poisson network with  $K$ replaced
by $K^{(\sigma)}$ given by~(\ref{e.Ksigma}).
\end{enumerate}
\end{remark}

Theorems~\ref{mainResult} and~\ref{mainResult2} also apply to shadowing models formed by a product of a log-normal and some other independent random variable, such as the Suzuki distribution, which seeks to capture both fast-fading and slow-shadowing. 
\begin{corollary}\label{mainCor2}
For iid positive random variables $F_i$ , if one replaces $S_i^{(\sigma)}$ with $S_i^{(\sigma)}F_i$,  then in the limit as $\sigma\rightarrow \infty$ the resulting propagation process also converges to an independently marked
 Poisson point process on $\R^+$ with intensity measure 
 \begin{equation} 
 \Lambda_F= \E(F^{2/\beta} ) \Lambda ,
  \end{equation}
  provided the moment condition
  \begin{equation} 
 \E(F^{2/\beta} )< \infty.
\end{equation}
  
\end{corollary}
\begin{IEEEproof}
\textcolor{blue}{Set $F_i$ as independent  marks and apply
Theorem~\ref{mainResult} given the values of $F_i$. We obtain the 
Poisson limit. Using the invariance property  
(Lemma~\ref{l.invariance}) the  limit remains Poisson, with an appropriately modified
mean, when $F_i$ are unconditioned.}
\end{IEEEproof}

Another obvious extension of the above convergence results exists for 
multi-tier network models where different stations of the network are
subject to different propagation loss conditions (fading distribution, path-loss
function). As the variance of the log-normally distributed shadowing
goes to infinity, different network tiers appear as independent
Poisson networks with possibly different parameters $a$.

\begin{remark}[Possible extensions of Theorem \ref{mainResult}]
\label{ss.future}
\textcolor{blue}{A natural question is whether the log-normal distribution of the shadowing is needed for the result to hold. 
The nature of our proof requires that the
random variables  $S_i$ can be written as the exponential of a
two-parameter infinitely divisible distribution.
However, as suggested in a recent \red{preprint}~\cite{Keeler-CSconvergence},
one can expect a similar Poisson \red{or, in certain cases, Cox convergence to occur} in a more general setting, when the shadowing, \red{fading or both} is constant in mean but converges in probability to zero. The new approach proposed in the aforementioned \red{preprint also leads to bounds} on
the distance between the given and asymptotic distribution, thus
allowing \red{one to quantify} the quality of approximation.}
 

\textcolor{red}{Another practical and no doubt challenging extension consists of studying cellular networks in the
presence of correlated shadowing.}

Finally, the majority of the research performed in this area has been under the assumption of a single typical user. An interesting task would be investigating this approach for two (or more) users, and hence deriving an equivalent version of the convergence result in this setting.
\end{remark}

\section{Numerical support for convergence results and a Linear-regression
estimation of path-loss exponent}
\label{s.Regression}

In this section we numerically verify the quality of the asymptotic
Poisson approximation regarding some simple model metrics
and present a new  linear-regression method for the estimation of the  path-loss exponent.

\subsection{Numerical support of Theorem~\ref{mainResult}}
To illustrate Theorem~\ref{mainResult} and obtain some insight into
the speed of convergence we used  Kolmogorov-Smirnov (K-S) test (\cite{Williams2001}) to compare the infinite Poisson model to the
hexagonal one regarding the cumulative distribution
functions (CDF) of the propagation loss from  the strongest base station
and the respective signal-to-interference-ratio (SIR).

\subsubsection{Smallest propagation loss}
Denote by 
$$L^*=\min\{L_i\in\Theta\}$$ 
the weakest propagation loss (usually
corresponding to the serving base station) in the Poisson
model considered in Section~\ref{s.Poisson}. Let us recall the
distribution of $L^*$.

\begin{corollary}
\label{f.MinCDF} Assume $\mathbf{E}\left[  S^{2/\beta}\right]  <\infty$.
Then in Poisson network with intensity $\lambda$ and arbitrary
distribution of $S$, we have
\begin{equation}
\mathbf{P}\left(  L^{\ast}\geq t\right)  =e^{-at^{2/\beta}}, \label{e.MinCDF}%
\end{equation}
where $a$ is given by Equation~(\ref{e.a}). In other words, the $1/L^{\ast}$
has a Fr\'{e}chet distribution with shape parameter $2/\beta$\ and scale
parameter $a^{\beta/2}$. \textcolor{blue}{(Equivalently, $L^{\ast}$ has a Weibull distribution.)} 
\end{corollary}

We will compare this distribution to the distribution of the weakest
propagation loss in a perfectly hexagonal network.
Specifically, we consider a hexagonal network $\phi_{H}^{N}$ of $N\times N$ base stations
located in the rectangle $[-N\Delta/2,N\Delta/2)\times\lbrack-N\sqrt{3}%
\Delta/4,N\sqrt{3}\Delta/4)$, where $\Delta$ is the distance between two
adjacent stations, and serving users located in this rectangle. Note that the
density of such a network is equal to $\lambda=2/(\Delta^{2}\sqrt{3})$. In
order to be able to neglect the boundary effects, let us assume the
\emph{toroidal metric} (\textquotedblleft wrap around\textquotedblright\ the
network, see~\cite{blaszczyszyn2010impact} for details).  We
consider also the log-normal shadowing as in the Poisson network and
for each its realization we place a user uniformly on the torus and  
find the weakest propagation loss $L^{\ast}$ measured 
with respect to any station in the toroidal network.
The closed form expression for the distribution of $L^*$
is not known,
hence we  simulate this  network and compare the {\em empirical CDF} of
$L^{\ast}$ to that given in Corollary~\ref{f.MinCDF} with the same
parameters (network density $\lambda$ and the shadowing parameter $\sigma$.) 

\begin{figure}[t]
\begin{center}
\includegraphics[width=0.5\linewidth]{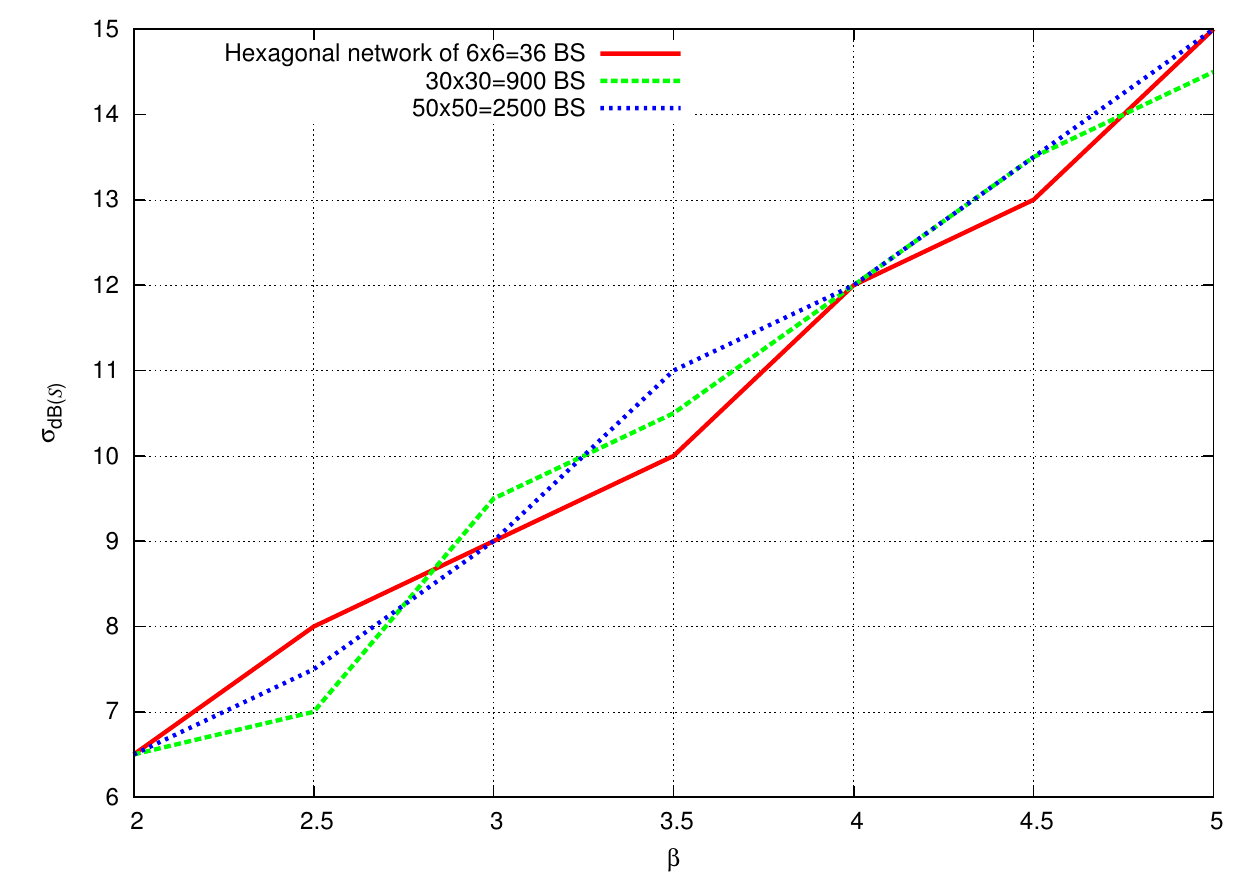}
\caption{\label{f.CSRLeft}Critical values of $\sigma_{\mathrm{dB}}$ in function of the path-loss exponent $\beta$, for different network sizes, above which, the empirical
distribution of $L^{\ast}$ in the hexagonal network cannot be distinguished
from this for the \textquotedblleft equivalent\textquotedblright\ Poisson
model at a 99\% confidence level, for 9/10 realizations of the hexagonal network. }%
\vspace{5ex}
\includegraphics[width=0.5\linewidth]{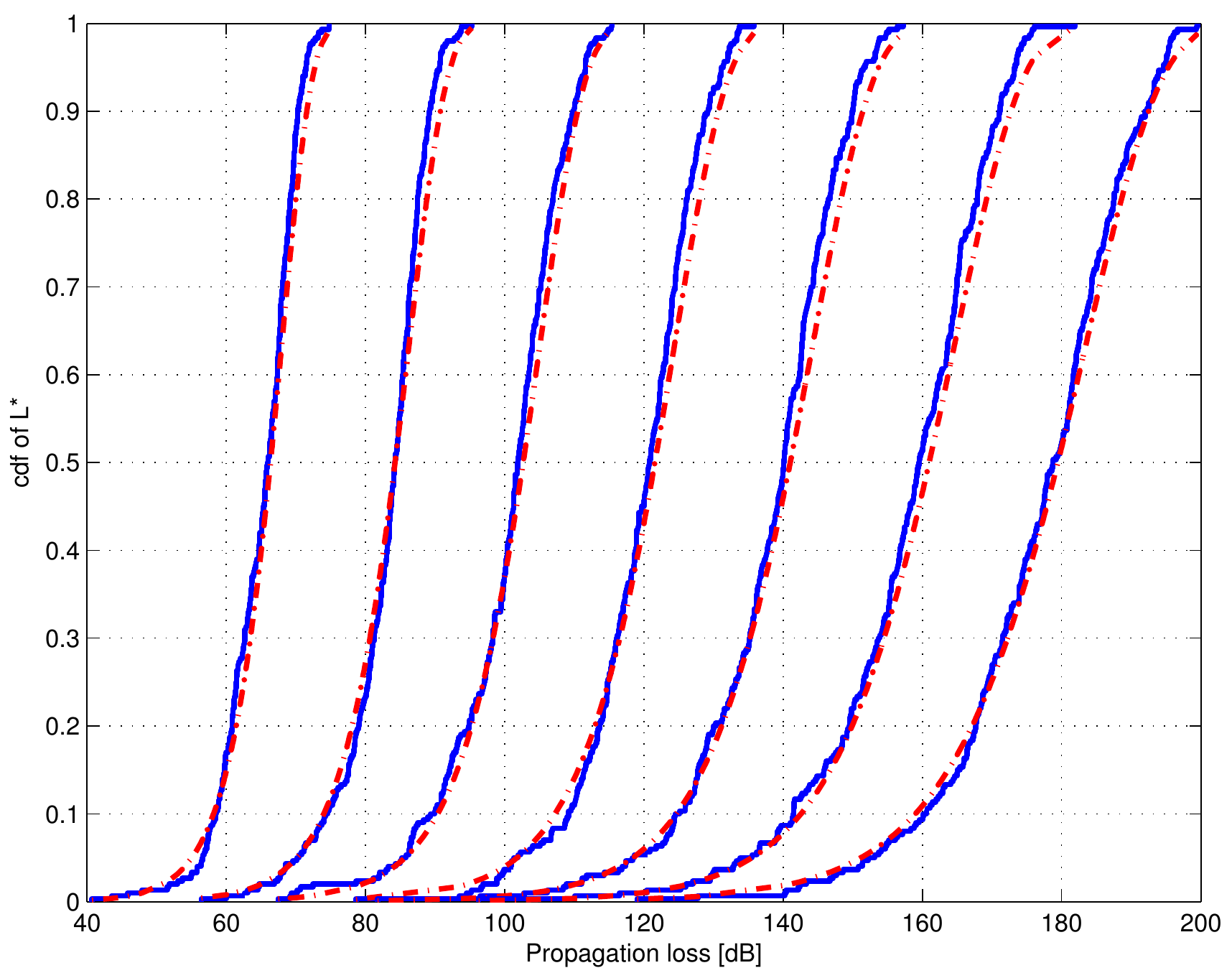}
\end{center}
\caption{\label{f.CSRRight} Visual comparison of the empirical distribution of $L^{\ast}$ in the
hexagonal network of $6\times6=36$ base stations (solid lines) to the theoretical cdf of
$L^{\ast}$ in the \textquotedblleft equivalent\textquotedblright\ Poisson
model (dashed lines) for $\beta=2,2.5,3,\ldots,5$ (curves from left to right)
and the corresponding critical values of $\sigma_{\mathrm{dB}}=\sigma
_{\mathrm{dB}}(\beta)$ taken from Figure~\ref{f.CSRLeft}. }%
\end{figure}

We observe that the \emph{supremum} (Kolmogorov) distance between the two
distributions decreases in $\sigma_{\mathrm{dB}}$. To make this observation
more quantitative, we perform a Kolmogorov-Smirnov (K-S) test (which is based
on this distance; cf.~\cite{Williams2001}) and we show in
Figure~\ref{f.CSRLeft} the values of $\sigma_{\mathrm{dB}}$, as a function of
$\beta$ for $N=6,30,50$, above which the K-S test does not allow one to one
distinguish the empirical distribution for the hexagonal model (based on 300
observations) from the closed-form (Poisson case) expression, at a 99\%
confidence level, for 9/10 realizations of the hexagonal network.
Figure~\ref{f.CSRRight} displays the goodness of fit for these critical values
of $\sigma_{\mathrm{dB}}:=\sigma 10/\log\!10\,\dB$\label{page.db}
\textcolor{blue}{(representing the standard deviation of the
  log-normal shadowing expressed in dB)} 
for $N=6$ (i.e.,
$6\times6=36$ base station network).

\paragraph{Signal to interference (SIR) distribution}
\begin{figure}[t!]
\begin{center}
\centerline{\includegraphics[width=0.5\linewidth]{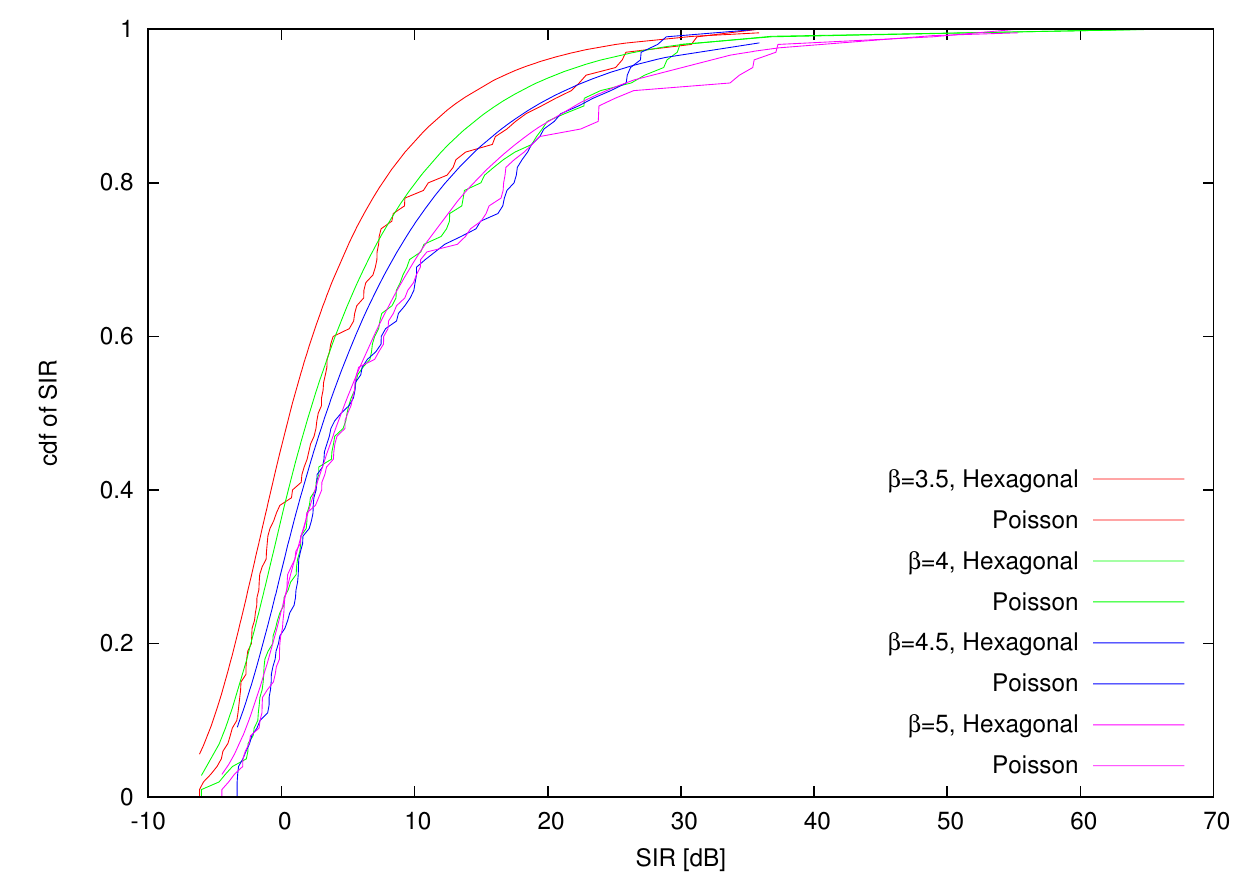}}
\caption{Empirical CDF of SIR simulated in hexagonal network with shadowing
and their Poisson approximations.
\label{f.SIR}}
\end{center}
\end{figure}

Consider again Poisson and  a hexagonal network  consisting of   $30\times30=900$
base stations  on a torus, with log-normal shadowing with the same
parameters. For these two networks we compare the distribution of the
SIR with respect to the strongest station
$$\text{SIR}=\frac{L^*}{\sum_{L_i\in\Theta}{L_i}-L^*}\,.$$
Recall that the analytic expression distribution of the SIR in Poisson
network is known; cf~\cite{kcovsingle,paul_matlab}.
In Figure~\ref{f.SIR} we present a few examples of these CDF's.
We found that for $9/10$ realizations of the network shadowing the K-S test does not allow one to distinguish the empirical (obtained from simulations)  CDF of the SIR  from the CDF of SIR evaluated in the infinite  Poisson model 
with the  critical $p$-value fixed to $\alpha=10\%$ provided  $\logsd$
is large enough. 

\subsection{Linear-regression estimation of parameters}

\label{s.Estimation} Based on the Poisson model, we will now suggest 
a method for the estimation of path-loss exponent $\beta$ from the measurements of the 
weakest propagation loss $L^{\ast}$ (performed by users in
operational networks and known by the network operator).  Corollary~\ref{f.MinCDF} implies
that
\begin{equation}
\log\left(  -\log\left[  \mathbf{P}\left(  L^{\ast}\geq t\right)  \right]
\right)  =\log a+\frac{2}{\beta}\log t\label{e.LinearFitting}%
\end{equation}
Consequently, if the distribution of $L^{\ast}$ is available from measurements
(or simulations), then one can estimate $\beta$ and $a$ by the linear regression
between $\log\left(  -\log\left[  \mathbf{P}\left(  L^{\ast}\geq t\right)
\right]  \right)  $ and~$\log t$. This characterizes in particular the
path-loss exponent $\beta$.

We will now apply this method with respect to data obtained from  simulations and
from measurements of $L^{\ast}$ realized and collected in the cellular network of
\emph{Orange} in a certain large city in Europe, with density $\lambda
=5.09\text{km}^{-2}$. For this dense urban area, we shall deduce the outdoor
and then the indoor path-loss exponent $\beta$.

\subsubsection{Outdoor}

The base stations positions are those of the UMTS network of \emph{Orange} in
a certain large city in Europe, operating with the carrier frequency
of~$2.1$GHz. The distribution of the outdoor propagation loss $L^{\ast}$ with
the serving base station is obtained by simulations performed with
\emph{StarWave}, a propagation software developed by \emph{Orange
Labs}~\footnote{This tool uses detailed information on the  terrain and buildings and
accounts for the diffraction, the guided propagation as well as the reflection
of the signal.}. The linear fitting~(\ref{e.LinearFitting}) of the empirical
data obtained from these simulations gives $\beta=3.85$;\footnote{The
95\%-confidence interval is $\beta\in\lbrack3.34,4.54]$; the Kolmogorov
distance between the empirical distribution and the estimated theoretical
distribution is $D=0.274$.}. This result can be validated by comparing it
to the value of $\beta$  given for example by the COST
Walfisch-Ikegami propagation model~\cite{Cost231_1999}. Indeed, 
considering  the COST Walfisch-Ikegami
model for the same frequency $2.1$GHz\footnote{The other network parameters
are: base station antenna height $30$m, mobile antenna height $1.5$m,
percentage of buildings $70$\%, nominal building height $25$m, building
separation $30$m and street width $20$m.} one obtains $\beta=3.80$ for
the non-line-of-sight propagation loss. Observe that the values of $\beta$ obtained
by the two different approaches are close to each other, which may validate
the novel approach for the data under consideration. 

\subsubsection{Indoor}

We now consider the actual users' data collected in the GSM network of
\emph{Orange} operating with a $1.8$GHz carrier frequency~\footnote{In fact
the measurements concern the network operating on two frequency bands
$1800$MHz\ and $900$MHz. Users connect first on $1800$MHz, and in case of a
problem switch to $900$MHz. Therefore the reported data may lead to an
underestimation of large values of the propagation loss.}. The operator
estimates that approximately $80\%$ of users are indoors and the remaining
$20\%$ outdoors. The linear fitting~(\ref{e.LinearFitting}) gives $\beta
=3.64$;\footnote{The 95\%-confidence interval is $\beta\in\lbrack3.42,3.88]$;
the Kolmogorov distance between the empirical distribution and the estimated
theoretical distribution is $D=0.119$. Note that this is a better fit than in
the case of the outdoor data.}. We are not aware of any alternative model
valid for indoor scenario; the COST Walfisch-Ikegami model being only valid
for outdoor scenario.

\section{Conclusion}
We presented a convergence result  to show that in the presence of
sufficiently large log-normal-based shadowing  the propagation processes (and, hence, SINR, spectral efficiency, etc,) experienced by a typical user in an empirically homogeneous wireless network behave stochastically as though the underlying base station configurations were scattered according to a homogeneous Poisson point process. In this shadowing regime, we also show that the distances to each base station are independent log-normal variables,  hence the propagation process has become decoupled from the underlying geometry. This decoupling carries a trade-off between exact geometric
information being lost and a considerable increase in tractability. Based on these findings, we presented a linear-regression method for estimating the exponent of the path-loss function based on user data in an empirically homogeneous cellular network with sufficiently large log-normal shadowing.  This novel method for estimating statistical parameters of the propagation model complements other models such as those of Hata or COST Walfisch-Ikegami. 

\appendix
\label{s.Appendix}
\subsection{Proof of Proposition~\ref{p.markedinvariance}}
\label{app.markedinvariance}
By the displacement theorem~\cite[Section 1.3.3]{FnT1} 
$\tilde\Theta$ is an independently  marked Poisson point process with intensity measure
\OneOrTwoColumnDisplay{
\begin{align*}
\tilde\Lambda((0,y)\times(0,\rho)\times\tau)
&=2\pi \lambda \E \int_0^{\infty}\Ind ( ( r K)^{\beta}/S \leq y ) \Ind (  r \leq \rho ) \Ind(T\in \tau) r\,dr\\
&=\pi \lambda \E [ \min(\rho^2,( y S)^{2/\beta}/K^2)  )  \Ind(T\in \tau) ]\\
&=\frac{\pi \lambda}{K^2} \E [S^{2/\beta}  \min((\rho K)^2/S^{2/\beta}, y^{2/\beta} )  \Ind(T \in \tau) ]\\
&=\frac{\pi \lambda}{K^2} \E [S^{2/\beta}  \min((\ell(\rho)/S)^{2/\beta}, y^{2/\beta} )  \Ind(T \in \tau) ]\\
&=\frac{2\pi \lambda}{\beta K^2} \E [S^{2/\beta} \int_0^y  \Ind (S \leq\ell(\rho)/u)  \Ind(T \in \tau) u^{2/\beta-1} \,du  ]\\
&=\frac{2 \pi \lambda \E(S^{2/\beta}) }{\beta K^2 \E(S^{2/\beta})} \int_0^y  \E [S^{2/\beta} \Ind (S \leq\ell(\rho)/u)  \Ind(T \in \tau)] u^{2/\beta-1}\,du\,, 
\end{align*}
}
{
\begin{align*}
&\tilde\Lambda((0,y)\times(0,\rho)\times\tau)\\
&=2\pi \lambda \E \int_0^{\infty}\Ind ( ( r K)^{\beta}/S \leq y ) \Ind (  r \leq \rho ) \Ind(T\in \tau) r\,dr\\
&=\pi \lambda \E [ \min(\rho^2,( y S)^{2/\beta}/K^2)  \Ind(T\in \tau) ]\\
&=\frac{\pi \lambda}{K^2} \E [S^{2/\beta}  \min((\rho K)^2/S^{2/\beta}, y^{2/\beta} )  \Ind(T \in \tau) ]\\
&=\frac{\pi \lambda}{K^2} \E [S^{2/\beta}  \min((\ell(\rho)/S)^{2/\beta}, y^{2/\beta} )  \Ind(T \in \tau) ]\\
&=\frac{2\pi \lambda}{\beta K^2} \E [S^{2/\beta} \int_0^y  \Ind (S \leq\ell(\rho)/u)  \Ind(T \in \tau) u^{2/\beta-1}\, du  ]\\
&=\frac{2 \pi \lambda \E(S^{2/\beta}) }{\beta K^2 \E(S^{2/\beta})} \int_0^y  \E [S^{2/\beta} \Ind (S \leq\ell(\rho)/u)  \Ind(T \in \tau)] u^{2/\beta-1} \,du\,,
\end{align*}
} 
where in the last line we exchanged the integral and expectation. Putting $\rho=\infty$ and $\tau=\calT$ allows one to  
recognize that
\begin{align*}
\Lambda(du)&=\frac{2\pi \lambda \E(S^{2/\beta})}{\beta K^2} u^{2/\beta-1} du\\
&=a \frac{2}{\beta}u^{2/\beta-1} du,
\end{align*}
is the intensity measure of the unmarked process $\Theta$.

\subsection{Proof of Proposition~\ref{p.LNGuT}}
\label{app.LNGuT}
Write $h_{\rho}:= (K\rho)^\beta/u$ and observe that 
\OneOrTwoColumnDisplay{
\begin{align*}
\E\Bigl[S^{2/\beta}\Ind (S   \leq h_{\rho})\Ind(T\in \tau)]) 
&=\E[S^{2/\beta}\Ind (S   \leq h_{\rho})G^{T|S}(\tau|S)\Bigr] \\
&=\frac{1}{\sqrt{2\pi}}\!\!\int_{-\infty}^{\infty}\hspace{-1em}e^{-\frac{t^2}{2}+\frac{2\sigma t+2\mu}{\beta} }\Ind (e^{\sigma t+\mu
}\leq h_{\rho}) G^{T|S}(\tau|e^{\sigma t+\mu})dt.
\end{align*}
}
{
\begin{align*}
\E&\Bigl[S^{2/\beta}\Ind (S   \leq h_{\rho})\Ind(T\in \tau)])  \\
&=\E[S^{2/\beta}\Ind (S   \leq h_{\rho})G^{T|S}(\tau|S)\Bigr] \\
&=\frac{1}{\sqrt{2\pi}}\!\!\int_{-\infty}^{\infty}\hspace{-1em}e^{-\frac{t^2}{2}+\frac{2\sigma t+2\mu}{\beta} }\Ind (e^{\sigma t+\mu
}\leq h_{\rho}) G^{T|S}(\tau|e^{\sigma t+\mu})dt.
\end{align*}
} 
Some algebra and~(\ref{e.S2/b}) gives
\OneOrTwoColumnDisplay{
\begin{align*}
\E[S^{2/\beta}\Ind (S   \leq h_{\rho})\Ind(T\in \tau)]/\E(S^{2/\beta}) 
&=
\frac{1}{\sqrt{2\pi}}\int_{-\infty}^{\infty}e^{-\frac{(t-2\sigma/\beta)^2}{2}} 
\Ind ( e^{\sigma t+\mu} \leq
h_{\rho})G^{T|S}(\tau|e^{\sigma t+\mu})  dt\\
&=\frac{1}{\sqrt{2\pi}}\int_{-\infty}^{\infty}e^{-\frac{t^2}{2}} 
\Ind ( e^{\sigma t+\mu+\frac{2\sigma^2}{\beta}} \leq
h_{\rho})G^{T|S}(\tau|e^{\sigma
  t+\mu+\frac{2\sigma^2}{\beta}})  dt\\
&=\E\Bigl[\Ind (Se^{2\sigma^2/\beta}  \leq (K\rho)^\beta/u)
G^{T|S}(\tau|Se^{2\sigma^2/\beta})\Bigr]\,.
\end{align*}
}
{
\begin{align*}
&\E[S^{2/\beta}\Ind (S   \leq h_{\rho})\Ind(T\in \tau)]/\E(S^{2/\beta}) \\
&=
\frac{1}{\sqrt{2\pi}}\int_{-\infty}^{\infty}e^{-\frac{(t-2\sigma/\beta)^2}{2}} 
\Ind ( e^{\sigma t+\mu} \leq
h_{\rho})G^{T|S}(\tau|e^{\sigma t+\mu})  dt\\
&=\frac{1}{\sqrt{2\pi}}\int_{-\infty}^{\infty}e^{-\frac{t^2}{2}} 
\Ind ( e^{\sigma t+\mu+\frac{2\sigma^2}{\beta}} \leq
h_{\rho})G^{T|S}(\tau|e^{\sigma
  t+\mu+\frac{2\sigma^2}{\beta}})  dt\\
&=\E\Bigl[\Ind (Se^{2\sigma^2/\beta}  \leq (K\rho)^\beta/u)
G^{T|S}(\tau|Se^{2\sigma^2/\beta})\Bigr]\,.
\end{align*}
} 
Equations (\ref{e.GT}) and (\ref{e.GR}) follow from~(\ref{e.Gu-LN})
by assuming $\rho=\infty$ and $\tau=\calT$, respectively.

\subsection{Proof of Theorem~\ref{mainResult}}
\label{AppMainTh}

We simplify (and slightly abuse) the notation by setting $n:=\sigma^2$
and write $n$ instead of $n^{1/2}$ in the subscripts and
superscripts. Let $n$ takes positive integer values without loss of
generality.  Furthermore, it is more convenient to study  the
propagation loss process $\markN$ on the logarithmic scale, hence we
denote by $\Lambda_{\log}$  the image of the 
measure $\Lambda$ given in Lemma~\ref{l.invariance}
 through the logarithmic mapping; i.e., 
\begin{equation}\label{e.Lambda}
\Lambda_{\log}((-\infty,s])
:= \int_{R^+}\Ind(\log(t)\le s)\,\Lambda(dt)=\frac{\lambda\pi}{K^2}\exp\left[\frac{2s}{\beta}\right]
\end{equation}
for $s\in\R$.

For all  $n\ge 1$ and $r\ge0$ we define and  observe
by~(\ref{e.lognormalRV}) and~(\ref{e.Ksigma}) that 
\begin{align}\nonumber
\nu_n (s,r)&:=\Prob \left[  \log\left(\frac{(K^{(n)})^{\beta} r^{\beta} }{S^{(n)}} \right) \leq
s\right ] \\
&=\Prob \left[Z \ge - 
\frac{s-\beta\log(Kr)-n/\beta}{\sqrt{n}} \right ]\nonumber \\
&=G_Z \left[\frac{s-\beta\log(Kr)-n/\beta}{\sqrt{n}} \right ],\label{v_nFinal}
\end{align}
where $G_Z$ is the CDF of the standard Gaussian random variable.

Let $B_0(r)=\{x\in\R^2:|x|<r\}$. 
We now need to derive two results.
\begin{lemma}\label{Lemma1}
\begin{align*}
\lim_{n\rightarrow \infty} \int_{B_0(b_n)\setminus B_0(a_n)}
\nu_n(s,|x|)\,\lambda dx &=\lim_{n\rightarrow
  \infty}\int_{\R^2}\nu_n(s,|x|)\,\lambda dx \\
& =\Lambda_{\log}((-\infty,s])
\end{align*}
provided that $a_n$ and $b_n$ satisfy (\ref{a_n}) and (\ref{b_n}).
\end{lemma}
\begin{IEEEproof}
We first examine the integral over $B_0(b_n)\setminus B_0(a_n)$  in polar coordinates
\OneOrTwoColumnDisplay{
\begin{equation}
\int_{B_0(b_n)\setminus B_0(a_n)}\nu_n (s,|x|)dx 
= 2\pi
\int_{a_n}^{b_n}rG_Z \left(\frac{s-\beta\log(Kr)-n/\beta}{\sqrt{n}} \right)dr.
\label{e.1}
\end{equation}
}
{
\begin{align}\nonumber 
&\int_{B_0(b_n)\setminus B_0(a_n)}\nu_n (s,|x|)dx \\
&= 2\pi
\int_{a_n}^{b_n}rG_Z \left(\frac{s-\beta\log(Kr)-n/\beta}{\sqrt{n}} \right)dr.
\label{e.1}
\end{align}
}  
The change of variables $t=(s-\beta\log(Kr)-n/\beta)/\sqrt{n}$ gives 
\begin{align}\nonumber
&2\pi
\int_{a_n}^{b_n}rG_Z\left(\frac{s-\beta\log(Kr)-n/\beta}{\sqrt{n}}
  \right )dr
\nonumber\\
&=2\pi \int_{v_n}^{u_n} \frac{1}{K^2}\exp\left[
\frac{2}{\beta}\left(s-t\sqrt{n} -n/\beta \right) \right] 
G_Z (t)\frac{\sqrt{n}}{\beta}dt \nonumber \\
&=2\pi\frac{\sqrt{n}}{\beta} \frac{\exp\left[ \frac{2}{\beta}\left(s
-\frac n\beta \right) \right]}{K^2}
 \int_{-u_n}^{-v_n} \exp\left[ \frac{2t\sqrt{n}}{\beta} \right]  {G_Z} (-t)\,dt 
\label{e.2}
\end{align}

where 
\begin{align}
&u_n=\frac{s-\beta\log(K a_n )-n/\beta}{\sqrt{n}},\label{e.un} \\
&v_n=\frac{s-\beta\log(Kb_n)-n/\beta}{\sqrt{n}}\,.\label{e.vn}
\end{align}
Moreover,
\OneOrTwoColumnDisplay{
\begin{align}
  \int_{-u_n}^{-v_n} \exp\left[ \frac{2t\sqrt{n}}{\beta} \right] 
{G_Z }(-t)dt 
=&\left.\frac{\beta}{2\sqrt{n}} \exp\left[ \frac{2t\sqrt{n}}{\beta} \right] 
{G_Z }(-t)\right| _{t=-u_n}^{t=-v_n} \label{e.31}\\
&+\frac{\beta}{2\sqrt{n}} \exp\left[\frac{2n}{\beta^2} \right]
 \int_{-u_n}^{-v_n} \exp\left[-\frac{1}{2} \left(t-\frac{2\sqrt{n}}{\beta}
\right)^2\right] \frac{dt}{\sqrt{2\pi}}.\label{e.32}
\end{align}
}
{
\begin{align}\nonumber
&  \int_{-u_n}^{-v_n} \exp\left[ \frac{2t\sqrt{n}}{\beta} \right] 
{G_Z }(-t)dt \nonumber\\ 
=&\left.\frac{\beta}{2\sqrt{n}} \exp\left[ \frac{2t\sqrt{n}}{\beta} \right] 
{G_Z }(-t)\right| _{t=-u_n}^{t=-v_n} \label{e.31}\\
&+\frac{\beta}{2\sqrt{n}} \exp\left[\frac{2n}{\beta^2} \right]
 \int_{-u_n}^{-v_n} \exp\left[-\frac{1}{2} \left(t-\frac{2\sqrt{n}}{\beta}
\right)^2\right] \frac{dt}{\sqrt{2\pi}}.\label{e.32}
\end{align}
}  
Combining~(\ref{e.1}), (\ref{e.2}) and (\ref{e.31})--(\ref{e.32}) we have
\OneOrTwoColumnDisplay{
\begin{align}
\int_{B_0(b_n)\setminus B_0(a_n)}\nu_n (s,|x|)dx  
=&\left.\frac{\pi}{K^2}\exp\left[ \frac{2}{\beta}\left(s
-\frac{n}{\beta}+t\sqrt n \right) \right]
{G_Z }(-t)\right| _{-u_n}^{-v_n} \label{e.integrated-term}\\
&+\frac{\pi}{K^2}\exp\left[ \frac{2s}{\beta} \right]
\int_{-u_n-\frac{2\sqrt{n}}{\beta}}^{-v_n-\frac{2\sqrt{n}}{\beta}} 
e^{-\frac{w^2}{2}} \frac{dw}{\sqrt{2\pi}},\label{e.integral}
\end{align}
}
{
\begin{align}
&\int_{B_0(b_n)\setminus B_0(a_n)}\nu_n (s,|x|)dx  \nonumber\\
=&\left.\frac{\pi}{K^2}\exp\left[ \frac{2}{\beta}\left(s
-\frac{n}{\beta}+t\sqrt n \right) \right]
{G_Z }(-t)\right| _{-u_n}^{-v_n} \label{e.integrated-term}\\
&+\frac{\pi}{K^2}\exp\left[ \frac{2s}{\beta} \right]
\int_{-u_n-\frac{2\sqrt{n}}{\beta}}^{-v_n-\frac{2\sqrt{n}}{\beta}} 
e^{-\frac{w^2}{2}} \frac{dw}{\sqrt{2\pi}},\label{e.integral}
\end{align}
}
where in the last integral we have changed the  variable $w=t-2
\sqrt{n}/\beta$.
Note first that the integrated term~(\ref{e.integrated-term})
is finite even if $v_n=-\infty$ (which is equivalent to $b_n=\infty$)
for some $n$, which can be deduced from the inequality
\begin{align}
G_Z(-t)=1-G_Z(t)
&=\frac{1}{\sqrt{2\pi}}\int_t^{\infty}e^{-x^2/2}dx
\le e^{-t^2/2}, \quad t\geq 0\,, \label{errIneq}
\end{align}
(see \cite[Section~7.8]{DLMF}).
By~(\ref{e.un})--(\ref{e.vn})  
\begin{align}
-u_n&=\sqrt n\left(\frac{\beta\log (a_n)}{n}+\frac{1}{\beta}\right)+o(1)
\label{e.-un}\\
-v_n&=\sqrt n\left(\frac{\beta\log (b_n)}{n}+\frac{1}{\beta}\right)+o(1)
\label{e.-vn}
\end{align}
as $n\to\infty$. Consequently, under conditions
(\ref{a_n}) and (\ref{b_n}) respectively, $-u_n-2\sqrt n/\beta\to
-\infty$ and
$-v_n-2\sqrt n/\beta\to \infty$, making the term~(\ref{e.integral})
converge to~$\lambda^{-1}\Lambda_{\log}((-\infty,s])$.

Regarding the integrated term~(\ref{e.integrated-term}), 
by~(\ref{b_n}) and~(\ref{e.-vn}),
$\log(b_n)/n\to\infty$ and $v_n\to-\infty$
by~(\ref{errIneq}) 
\OneOrTwoColumnDisplay{
\begin{align*}
\exp\left[\frac{2}{\beta}\left(s-\frac{n}{\beta}-v_n\sqrt{n}\right)\right]{G_Z }(v_n)\nonumber
&\le\exp\left[\frac{2}{\beta}\left(s-\frac{n}{\beta}\right)
+\frac{v_n}{2}\left(-v_n-\frac{4\sqrt n}{\beta}\right)\right]
\nonumber\\
&=
\exp\left[\frac{2}{\beta}\left(s-\frac{n}{\beta}\right)+
\frac{\sqrt n v_n}{2}\left(\frac{\beta\log(b_n)}{n}
-\frac{3}{\beta}+o(\frac{1}{\sqrt n})\right)\right]\nonumber\\
&\to 0\qquad (n\to\infty). 
\end{align*}
}
{
\begin{align*}
&\exp\left[\frac{2}{\beta}\left(s-\frac{n}{\beta}-v_n\sqrt{n}\right)\right]{G_Z }(v_n)\nonumber\\
&\le\exp\left[\frac{2}{\beta}\left(s-\frac{n}{\beta}\right)
+\frac{v_n}{2}\left(-v_n-\frac{4\sqrt n}{\beta}\right)\right]
\nonumber\\
&=
\exp\left[\frac{2}{\beta}\left(s-\frac{n}{\beta}\right)+
\frac{\sqrt n v_n}{2}\left(\frac{\beta\log(b_n)}{n}
-\frac{3}{\beta}+o(\frac{1}{\sqrt n})\right)\right]\nonumber\\
&\to 0\qquad (n\to\infty). 
\end{align*}
}  

Considering the term containing $-u_n$ in~(\ref{e.integrated-term}),
it is easy to see that it converges to~0 when $-u_n\le0$ (by the
trivial bound $G_Z(-t)\le1$). We will thus consider from now only $-u_n\ge0$.
Moreover by ~(\ref{a_n}) and~(\ref{e.un}) we have also  $-u_n\le \sqrt
n\Bigl(1/\beta+o(1)\Bigr)$. For such values $-u_n$
we use again~(\ref{errIneq}) and  observing that 
the function  $-2u_n\sqrt{n}/\beta-{u^2_n}/{2}={u_n}/{2}\left(-u_n-{4\sqrt n}/{\beta}\right)$
is increasing in $-u_n$ for $-u_n \le 2\sqrt n/\beta$ 
we obtain
\OneOrTwoColumnDisplay{
\begin{align*}
\exp\left[\frac{2}{\beta}\left(s-\frac{n}{\beta}-u_n\sqrt{n}\right)\right]{G_Z }(u_n)\nonumber
&\le\exp\left[\frac{2}{\beta}\left(s-\frac{n}{\beta}-u_n\sqrt{n}\right)
-\frac{u^2_n}{2}\right]
\nonumber\\
&\le\exp\left[\frac{2}{\beta}\left(s-\frac{n}{\beta}\right)
+\frac{u_n}{2}\left(-u_n-\frac{4\sqrt n}{\beta}\right)\right]
\nonumber\\
&\le
\exp\left[\frac{2}{\beta}\left(s-\frac{n}{\beta}
+n\Bigl(\frac{1}{\beta}+o(1)\Bigr)\right)
-\frac{n}{2}\Bigl(\frac{1}{\beta}+o(1)\Bigr)^2\right]\\
&=
\exp\left[\frac{2}{\beta}\left(s
-n\Bigl(\frac{1}{4\beta}+o(1)\Bigr)\right)\right]\to0\quad (n\to\infty),
\end{align*}
}
{
\begin{align*}
&\exp\left[\frac{2}{\beta}\left(s-\frac{n}{\beta}-u_n\sqrt{n}\right)\right]{G_Z }(u_n)\nonumber\\
&\le\exp\left[\frac{2}{\beta}\left(s-\frac{n}{\beta}-u_n\sqrt{n}\right)
-\frac{u^2_n}{2}\right]
\nonumber\\
&\le\exp\left[\frac{2}{\beta}\left(s-\frac{n}{\beta}\right)
+\frac{u_n}{2}\left(-u_n-\frac{4\sqrt n}{\beta}\right)\right]
\nonumber\\
&\le
\exp\left[\frac{2}{\beta}\left(s-\frac{n}{\beta}
+n\Bigl(\frac{1}{\beta}+o(1)\Bigr)\right)
-\frac{n}{2}\Bigl(\frac{1}{\beta}+o(1)\Bigr)^2\right]\\
&=
\exp\left[\frac{2}{\beta}\left(s
-n\Bigl(\frac{1}{4\beta}+o(1)\Bigr)\right)\right]\to0\quad (n\to\infty),
\end{align*}
}  
 which concludes the proof of Lemma~\ref{Lemma1}.
\end{IEEEproof}

\begin{lemma}\label{Lemma2}
Assume (\ref{lambdaDef}), (\ref{a_n}) and (\ref{b_n}), then
\OneOrTwoColumnDisplay{
\begin{align}
\lim_{n\rightarrow \infty} \sum_{X_i\in \phi\cap (B_0(b_n)\setminus B_0(a_n))}
\nu_n (s,|X_i|) 
&=\lim_{n\rightarrow \infty} \sum_{X_i\in \phi} \nu_n
(s,|X_i|)=
\Lambda_{\log}((-\infty,s]).\label{Lemma2eq2}
\end{align}
}
{
\begin{align}\label{Lemma2eq1}
&\lim_{n\rightarrow \infty} \sum_{X_i\in \phi\cap (B_0(b_n)\setminus B_0(a_n))}
\nu_n (s,|X_i|) \\
&=\lim_{n\rightarrow \infty} \sum_{X_i\in \phi} \nu_n
(s,|X_i|)=
\Lambda_{\log}((-\infty,s]).\label{Lemma2eq2}
\end{align}
}  
\end{lemma}

\begin{IEEEproof}
For $k\geq 0$ and a fixed $\epsilon>0$, let $r_k=e^{\epsilon k}$ and
$A_k=B_0(r_{k+1})\setminus B_0(r_k)$, and write the summation
in~(\ref{Lemma2eq2}) as  
\OneOrTwoColumnDisplay{
\begin{align}
\sum_{X_i\in \phi} \nu_n (s,|X_i|)
&=\sum_{X_i\in \phi\cap B_0(r_{k_0})} \nu_n
(s,|X_i|) + 
\sum_{k=k_0}^{\infty}\sum_{X_i\in \phi\cap A_k} \nu_n (s,|X_i|) ,
\label{e.2summands}
\end{align}
}
{
\begin{align}\nonumber
&\sum_{X_i\in \phi} \nu_n (s,|X_i|)\\
&=\sum_{X_i\in \phi\cap B_0(r_{k_0})} \nu_n
(s,|X_i|) + 
\sum_{k=k_0}^{\infty}\sum_{X_i\in \phi\cap A_k} \nu_n (s,|X_i|) ,
\label{e.2summands}
\end{align}
}  
for some $k_0 \geq 0$, whose value will be fixed later on. In the
limit when $n\rightarrow\infty$, the first summation
in~(\ref{e.2summands}) vanishes; indeed
\OneOrTwoColumnDisplay{
\begin{align*}
\sum_{X_i\in \phi\cap B_0(r_{k_0})} \nu_n (s,|X_i|)  
 &=  \sum_{X_i\in \phi\cap
B_0(r_{k_0})}\Prob \left[Z \leq 
\frac{s-\beta\log(K|X_i|)-n/\beta}{\sqrt{n}} \right ]\\
&=  \sum_{X_i\in \phi\cap B_0(r_{k_0})}
G_Z \left(\frac{s-\beta\log(K|X_i|)-n/\beta}{\sqrt{n}}
\right )\\
& \leq \phi(B_0(r_{k_0}))G_Z
\left(\frac{s-\beta\log(K|X_*|)-n/\beta}{\sqrt{n}} \right )
\rightarrow 0 \quad (n \rightarrow \infty),
\end{align*}
}
{
\begin{align*}
&\sum_{X_i\in \phi\cap B_0(r_{k_0})} \nu_n (s,|X_i|)  \\
 &=  \sum_{X_i\in \phi\cap
B_0(r_{k_0})}\Prob \left[Z \leq 
\frac{s-\beta\log(K|X_i|)-n/\beta}{\sqrt{n}} \right ]\\
&=  \sum_{X_i\in \phi\cap B_0(r_{k_0})}
G_Z \left(\frac{s-\beta\log(K|X_i|)-n/\beta}{\sqrt{n}}
\right )\\
& \leq \phi(B_0(r_{k_0}))G_Z
\left(\frac{s-\beta\log(K|X_*|)-n/\beta}{\sqrt{n}} \right )
\rightarrow 0 \quad (n \rightarrow \infty),
\end{align*}
}  
where $X_*$ gives the maximum of $G_Z
\left(\frac{s-\beta\log(K|X|)-n/\beta}{\sqrt{n}} \right )$ over
$X\in\phi\cap B_0(r_{k_0})$  which exists since $\phi$ is (by
our assumption) a locally finite point measure. 
For the second summation in~(\ref{e.2summands}) we write $\nu_n (s,|X_i|)=\nu_n
(s,|x|\frac{|X_i|}{|x|})$, hence
\begin{align*}
 \nu_n (s,|X_i|)
=\frac{1}{|A_k|}\int_{A_k}\nu_n \left(s,|x|\frac{|X_i|}{|x|}\right) dx.
\end{align*}
Then the bounds for $x,X_i\in A_k$
$$
e^{-\epsilon}=\frac{r_{k}}{r_{k+1}}\leq\frac{|X_i|}{|x|} \leq
\frac{r_{k+1}}{r_{k}}=e^{\epsilon},
$$
and the expression~(\ref{v_nFinal}) of $\nu_n$, which implies $\nu_n (s,|x|e^{\epsilon})=\nu_n
(s-\beta\epsilon,|x|)$, lead to the lower bound
\begin{equation}
\sum_{k=k_0}^{\infty}\sum_{X_i\in \phi\cap A_k} \nu_n (s,|X_i|) 
\geq
\sum_{k=k_0}^{\infty} \frac{\phi( A_k)}{|A_k|}\int_{A_k}\nu_n
(s-\beta\epsilon ,|x|) dx,
\label{lowerBound1}   
\end{equation}
and the upper bound
\begin{equation}
\sum_{k=k_0}^{\infty}\sum_{X_i\in \phi\cap A_k} \nu_n (s,|X_i|)  \\
\leq  \sum_{k=k_0}^{\infty} \frac{\phi( A_k)}{|A_k|}\int_{A_k}\nu_n
(s+\beta\epsilon,|x|) dx. \label{upperBound1}
\end{equation}
Moreover, we write
\begin{align*} 
\frac{\phi( A_k)}{|A_k|} &= \frac{\phi( B_0(r_{k+1}))-\phi(
B_0(r_{k}))}{|B_0(r_{k+1})|-|B_0(r_{k})|}\\
&= \frac{\frac{\phi( B_0(r_{k+1}))}{|B_0(r_{k+1})|}-\frac{\phi(
B_0(r_{k}))}{|B_0(r_{k})|}\frac{|B_0(r_{k})|}{|B_0(r_{k+1})|}}{1-\frac{|B_0(r_{k
})|}{|B_0(r_{k+1})|}}\\
&= \frac{\frac{\phi( B_0(r_{k+1}))}{|B_0(r_{k+1})|}-\frac{\phi(
B_0(r_{k}))}{|B_0(r_{k})|}e^{-2\epsilon}}{1-e^{-2\epsilon}},
\end{align*}
and the requirement (\ref{lambdaDef}) yields
$
\lim_{k\rightarrow \infty}\frac{\phi( A_k)}{|A_k|} = \lambda
$.
Hence, for any fixed $\delta>0$, there exists a $k_0(\delta)$ such that for
all $k\geq k_0$, the bounds
$$
(1-\delta)\lambda \leq \frac{\phi( A_k)}{|A_k|} \leq (1+\delta)\lambda,
$$
hold. Lower bound (\ref{lowerBound1}) becomes
\begin{align*} 
\sum_{k=k_0}^{\infty}\sum_{X_i\in \phi\cap A_k} \!\!\!\nu_n (s,|X_i|)
&\geq
\sum_{k=k_0}^{\infty} (1-\delta)\lambda \int_{A_k}\!\!\!\nu_n (s-\beta\epsilon,|x|)
dx,\\  
&=(1-\delta)\lambda \int_{|x|\geq r_{k_0}}\!\!\!\!\!\nu_n (s-\beta\epsilon,|x|) dx\,.
\end{align*}
Finally, Lemma \ref{Lemma1} allows us to set $a_n=r_{k_0}$ and $b_n=\infty$, hence 
$$
 \lim_{n\rightarrow \infty}\sum_{k=k_0}^{\infty}\sum_{X_i\in \phi\cap A_k}
\nu_n (s,|X_i|)\geq (1-\delta) \frac{\pi\lambda
 }{K^2} \exp\left[\frac{2(s-\beta\epsilon)}{\beta}\right],
$$
and similarly the upper bound (\ref{upperBound1}) becomes
$$ 
\lim_{n\rightarrow \infty}\sum_{k=k_0}^{\infty}\sum_{X_i\in \phi\cap A_k}
\nu_n (s,|X_i|)\leq (1+\delta) \frac{\pi\lambda
 }{K^2} \exp\left[\frac{2(s+\beta\epsilon)}{\beta}\right]\,.
$$
Letting $\epsilon\rightarrow 0$ and $\delta \rightarrow 0$
completes the proof of (\ref{Lemma2eq2}).
The other result, (\ref{Lemma2eq2}), can be proved by a
straightforward modification
of the above arguments.  
\end{IEEEproof}

\begin{IEEEproof}[Proof of Theorem \ref{mainResult}]
\textcolor{blue}{The classical convergence result~\cite[Theorem 11.2.V]{daleyPPII2008} 
in conjunction with \cite[(11.4.2) and (11.4.3)]{daleyPPII2008}
says that the propagation loss process $\widetilde \Theta$ converges to the Poisson limit provided}
\begin{equation}\label{cond1}
\sup_i \nu^i_n (A)  \rightarrow 0\quad (n \rightarrow \infty).
\end{equation}
and
\begin{equation}\label{cond2}
\sum_i \nu^i_n (A) \rightarrow \Lambda_{\log}(A) \quad
(n \rightarrow \infty),
\end{equation}
for all bounded Borel sets $A\subset\R$, 
where $\nu_n^i(\cdot)$  is the (probability) measure on $\R$ defined
by setting $\nu_n^i((-\infty,s]):=\nu_n(s,|X_i|)$.
The first condition, (\ref{cond1}), clearly holds by~(\ref{v_nFinal}) for any locally finite
$\phi$ without a point at the origin. 
The second condition, (\ref{cond2})
follows from Lemma~
\ref{Lemma2}, which establish the
required convergence for $A=(-\infty,s]$ and any~$s\in\R$. This is
enough to conclude the convergence for all bounded Borel sets.
\end{IEEEproof}

\subsection{Proof of Theorem~\ref{mainResult2}}
\label{AppMainTh2}
We begin by explaining the pertinence of our assumptions~(\ref{e.tRsigma}) and~(\ref{e.GTsigma}).
Let $S^{(\sigma)}$ be given by~(\ref{e.lognormalRV}), with $\mu=-\sigma^2/2$, and denote by $G^{T|Z}(\tau|z)$,  $z\in\ir$, $\tau\in\calT$, the conditional distribution of the mark $T$ given $Z$
(which is, recall, a standard Gaussian random variable)
\begin{equation}\label{e.T|Zsigma}
G^{T^{(\sigma)}|Z}(\tau|z):=G^{T^{(\sigma)}|S^{(\sigma)}}(\tau|e^{-\sigma^2/2+\sigma z})\,.
\end{equation}
Then a simple algebra allows one to express $G_u (\rho,\tau)$ given by~(\ref{e.Gu-LN}), with  $K^{(\sigma)}$ given by~(\ref{e.Ksigma})
and $S^{(\sigma)}$ as above in the following way
\OneOrTwoColumnDisplay{
\begin{align}\label{e.GurhoZ}
G_u (\rho,\tau)
= \E\left[\Ind\left(Z\le \frac{\beta\log
      \rho}{\sigma}-\frac{\sigma}{\beta}-\frac{\beta\log k -\log
      u}{\sigma}\right)
G^{T^{(\sigma)}|Z}\left(\tau\Big |Z+\frac{2\sigma}{\beta}\right)\right]\,.
\end{align}
}
{
\begin{align}\label{e.GurhoZ}
&G_u (\rho,\tau)\\ \nonumber
&= \E\left[\Ind\left(Z\le \frac{\beta\log
      \rho}{\sigma}-\frac{\sigma}{\beta}-\frac{\beta\log k -\log
      u}{\sigma}\right)\right.\\
&\hspace{0.5\linewidth}\times\left.
G^{T^{(\sigma)}|Z}\left(\tau\Big |Z+\frac{2\sigma}{\beta}\right)\right]\,.\nonumber
\end{align}
}  
Note that $(\beta\log k -\log u)/\sigma$  vanishes when  $\sigma\to\infty$ making the conditional distribution  $G_u (\rho,\tau)$
(of the distance and type of the base station whose propagation loss is equal to $u$) independent of $u$. Furthermore, (\ref{e.GurhoZ})
suggests~(\ref{e.tRsigma}) as the scaling of the distance to the base station
and (\ref{e.GTsigma}) for the  conditional distribution of the type $T_i^{(\sigma)}$ of the base station given the shadowing $S^{(\sigma)}$.

We proceed now to the proof of Theorem~\ref{mainResult2}.
In analogy to~(\ref{e.Lambda}) we define 
the image of the measure $\tilde\Lambda$ given in Theorem~\ref{mainResult2} through the
logarithmic mapping of the propagation loss values
$$
\tilde\Lambda_{\log}((-\infty,s]\times(-\infty,\rho]\times\tau)
=\frac{\lambda\pi}{K^2}\exp\left[\frac{2s}{\beta}\right]G(\rho;\tau)$$
for $s\in\R$, $\rho\in\ir$ and $\tau\in\calT$.
Similarly, we extend the measures~(\ref{v_nFinal}) for $n\ge1$, $r,\rho\ge0$ and $\tau\in\calT$
and observe that, by~(\ref{e.tRsigma}) and~(\ref{e.GTsigma}) 
\OneOrTwoColumnDisplay{
\begin{align}\nonumber
\tilde\nu_n (s,\rho,\tau,r)
&:=\Prob \left[  \log\left(\frac{(K^{(n)})^{\beta}r^{\beta} }{\exp[-n/2+\sqrt nZ]} \right) \leq s, \calR(r)\le\rho,T^{(n)}\in\tau\right ] \nonumber\\[1ex]
&=\Prob\left(Z\ge -\frac{s-\beta\log(Kr)-n/\beta}{\sqrt{n}},T^{(n)}\in\tau \right)\ind(\calR(r)\le\rho)\\
&=G^* \left(-\frac{s-\beta\log(Kr)-n/\beta}{\sqrt{n}}; \tau\right)\ind(\calR(r)\le\rho)\,,\label{v_nFinal-tilde}
\end{align}
}
{
\begin{align}\nonumber
&\tilde\nu_n (s,\rho,\tau,r)\\
&:=\Prob \left[  \log\left(\frac{(K^{(n)})^{\beta}r^{\beta} }{\exp[-n/2+\sqrt nZ]} \right) \leq s, \calR(r)\le\rho,T^{(n)}\in\tau\right ] \nonumber\\[1ex]
&=\Prob\left(Z\ge -\frac{s-\beta\log(Kr)-n/\beta}{\sqrt{n}},T^{(n)}\in\tau \right)\ind(\calR(r)\le\rho)\\
&=G^* \left(-\frac{s-\beta\log(Kr)-n/\beta}{\sqrt{n}}; \tau\right)\ind(\calR(r)\le\rho)\,,\label{v_nFinal-tilde}
\end{align}
}  
where 
$$G^*(z;\tau)=\int_z^\infty G^{T|Z}(\tau|z-2\sqrt n/\beta)\,G_Z(dz)\,.$$
With this notation we extend Lemma~\ref{Lemma1} (cf its proof for the details and $u_n$, $v_n$) 
\OneOrTwoColumnDisplay{
\begin{align}
\int_{B_0(b_n)\setminus B_0(a_n)} \tilde\nu_n(s,\rho,\tau,|x|)dx 
&= 2\pi\!\!
\int_{a_n}^{b_n}\hspace{-0.8em}rG^* \left(-\frac{s-\beta\log(Kr)-n/\beta}{\sqrt{n}};\tau \right)
\!\!\ind\!\!\left(r\le e^{\frac{\sqrt n}{\beta}(\rho+\frac{\sqrt n}{\beta})}\right)dr \nonumber\\
&=2\pi\frac{\sqrt{n}}{\beta} \frac{e^{\frac{2}{\beta}\left(s
-n/\beta \right)}}{K^2}
\hspace{-0.5em} \int_{-u_n}^{-v_n}\hspace{-1em}
e^{\frac{2t\sqrt{n}}{\beta}} {\scriptstyle {G^*} (t;\tau)
\ind\left(t\le \rho+\frac{2\sqrt n}{\beta}-\frac{s-\beta\log K}{\sqrt n}\right)}dt \nonumber\\
&=\left.\frac{\pi}{K^2}e^{\frac{2}{\beta}\left(s
-\frac{n}{\beta}+t\sqrt n \right)}
{G^* }(t;\tau){\scriptstyle\ind\left(t\le \rho+\frac{2\sqrt n}{\beta}-\frac{s-\beta\log K}{\sqrt n}\right)}
\right| _{-u_n}^{-v_n} \label{e.integrated-term-tilde} \\
&\hspace{1em}+\frac{\pi}{K^2}e^{\frac{2s}{\beta}}
\int_{-u_n-\frac{2\sqrt{n}}{\beta}}^{-v_n-\frac{2\sqrt{n}}{\beta}} 
e^{-\frac{w^2}{2}}
{\scriptstyle \ind\left(w\le \rho-\frac{s-\beta\log K}{\sqrt n}\right)}G^{T|Z}(\tau|w)
 \frac{dw}{\sqrt{2\pi}}\nonumber\\
&\hspace{8em}\to\frac{\pi}{K^2} e^{\frac{2s}{\beta}}G(\rho;\tau)\qquad (n\to\infty),
\nonumber 
\end{align}
}
{
\begin{align}
&\int_{B_0(b_n)\setminus B_0(a_n)} \tilde\nu_n(s,\rho,\tau,|x|)dx \nonumber\\
&= 2\pi\!\!
\int_{a_n}^{b_n}\hspace{-0.8em}rG^* \left(-\frac{s-\beta\log(Kr)-n/\beta}{\sqrt{n}};\tau \right)
\!\!\ind\!\!\left(r\le e^{\frac{\sqrt n}{\beta}(\rho+\frac{\sqrt n}{\beta})}\right)dr \nonumber\\
&=2\pi\frac{\sqrt{n}}{\beta} \frac{e^{\frac{2}{\beta}\left(s
-n/\beta \right)}}{K^2}
\hspace{-0.5em} \int_{-u_n}^{-v_n}\hspace{-1em}
e^{\frac{2t\sqrt{n}}{\beta}} {\scriptstyle {G^*} (t;\tau)
\ind\left(t\le \rho+\frac{2\sqrt n}{\beta}-\frac{s-\beta\log K}{\sqrt n}\right)}dt \nonumber\\
&=\left.\frac{\pi}{K^2}e^{\frac{2}{\beta}\left(s
-\frac{n}{\beta}+t\sqrt n \right)}
{G^* }(t;\tau){\scriptstyle\ind\left(t\le \rho+\frac{2\sqrt n}{\beta}-\frac{s-\beta\log K}{\sqrt n}\right)}
\right| _{-u_n}^{-v_n} \label{e.integrated-term-tilde} \\
&\hspace{1em}+\frac{\pi}{K^2}e^{\frac{2s}{\beta}}
\int_{-u_n-\frac{2\sqrt{n}}{\beta}}^{-v_n-\frac{2\sqrt{n}}{\beta}} 
e^{-\frac{w^2}{2}}
{\scriptstyle \ind\left(w\le \rho-\frac{s-\beta\log K}{\sqrt n}\right)}G^{T|Z}(\tau|w)
 \frac{dw}{\sqrt{2\pi}}\nonumber\\
&\hspace{8em}\to\frac{\pi}{K^2} e^{\frac{2s}{\beta}}G(\rho;\tau)\qquad (n\to\infty),
\nonumber 
\end{align}
}  
because $0\le G^*(t;\tau)\le 1-G_Z(t)=G_Z(-t)$ thus the 
 term~(\ref{e.integrated-term-tilde}), being dominated by~(\ref{e.integrated-term}),
converges to~0.
Moreover, using the same arguments as in the proof  Lemma~\ref{Lemma2},
with $\tilde\nu_n (s,\rho,\tau,re^\epsilon)=\tilde\nu_n (s-\beta\epsilon,\rho-\beta\epsilon/\sqrt n,\tau,r)$, one proves
\begin{align*}
&\lim_{n\rightarrow \infty} \sum_{X_i\in \phi\cap (B_0(b_n)\setminus B_0(a_n))}
\tilde\nu_n (s,\rho,\tau,|X_i|) \\
&=\lim_{n\rightarrow \infty} \sum_{X_i\in \phi} \tilde\nu_n
(s,\rho,\tau,|X_i|)=
\frac{\lambda\pi}{K^2} e^{\frac{2s}{\beta}}G(\rho;\tau)\,.
\end{align*}
Now, the result follows by the same arguments as used in the proof
 of Theorem \ref{mainResult}.

\addtocounter{section}{1}
\addcontentsline{toc}{section}{References} 
\bibliographystyle{IEEEtran}

\end{document}